\documentclass[structabstract]{aa}  
\usepackage{graphicx}
\usepackage{txfonts}
\usepackage{float}
\usepackage[]{natbib}
\newcommand{\ud}{\mathrm{d}}

\begin{document}
   \title{Laboratory demonstration of a mid-infrared \\AGPM vector vortex coronagraph}

\author{C.~Delacroix\inst{1} 
\and O.~Absil\inst{1} 
\and P.~Forsberg\inst{2}
\and D.~Mawet\inst{3}
\and V.~Christiaens\inst{1}
\and M.~Karlsson\inst{2}
\and \\A.~Boccaletti\inst{4}
\and P.~Baudoz\inst{4}
\and M.~Kuittinen\inst{5}
\and I.~Vartiainen\inst{6} 
\and J.~Surdej\inst{1}
\and S.~Habraken\inst{1}
}

\institute{D\'epartement d'Astrophysique, G\'eophysique et Oc\'eanographie, Universit\'e de Li\`ege, All\'ee du Six Ao\^ut 17, B-4000 Li\`ege, Belgique
\and Department of Engineering Sciences, {\AA}ngstr\"{o}m Laboratory, Uppsala University, L\"{a}gerhyddsv\"{a}gen 1, SE-751 21 Uppsala, Sweden
\and European Southern Observatory, Alonso de Cordova 3107, Casilla 19001, Vitacura, Santiago 19, Chile
\and LESIA-Observatoire de Paris, CNRS, UPMC Univ.\ Paris 06, Univ.\ Paris-Diderot, 5 pl.~J.~Janssen, F-92195 Meudon, France
\and Department of Physics and Mathematics, University of Eastern Finland, P.O.~Box 111, FI-80101 Joensuu, Finland
\and Paul Scherrer Institut, 5232 Villigen PSI, Switzerland
}

\date{Received 18 January 2013 / Accepted 2 April 2013}

 
\abstract
{Coronagraphy is a powerful technique to achieve high contrast imaging and hence to image faint companions around bright targets. Various concepts have been used in the visible and near-infrared regimes, while coronagraphic applications in the mid-infrared remain nowadays largely unexplored. Vector vortex phase masks based on concentric subwavelength gratings show great promise for such applications.}
{We aim at producing and validating the first high-performance broadband focal plane phase mask coronagraphs for applications in the mid-infrared regime, and in particular the L band with a fractional bandwidth of $\sim$16\% (3.5-4.1 $\mu$m).}
{Based on rigorous coupled wave analysis, we designed an annular groove phase mask (AGPM) producing a vortex effect in the L band, and etched it onto a series of diamond substrates. The grating parameters were measured by means of scanning electron microscopy. The resulting components were then tested on a mid-infrared coronagraphic test bench.}
{A broadband raw null depth of $2\times 10^{-3}$ was obtained for our best L-band AGPM after only a few iterations between design and manufacturing. This corresponds to a raw contrast of about $6\times10^{-5}$ (10.5 mag) at $2\lambda/D$. This result is fully in line with our projections based on rigorous coupled wave analysis modeling, using the measured grating parameters. The sensitivity to tilt and focus has also been evaluated.}
{After years of technological developments, mid-infrared vector vortex coronagraphs finally become a reality and live up to our expectations. Based on their measured performance, our L-band AGPMs are now ready to open a new parameter space in exoplanet imaging at major ground-based observatories.}


\maketitle


\section{Introduction} \label{sec:intro}

Direct imaging of exoplanets and close environments of stars has recently made a long-awaited breakthrough \citep{Marois2008, Kalas2008, Lagrange2009a, Lagrange2010, AbsilMawet2010}. While coronagraphs have sometimes been parts of these discoveries, their role was restricted to merely mitigate detector saturation. The reason for the still limited impact of stellar coronagraphs in direct imaging is twofold. First, the wavefront quality and stability provided by current instruments on most ground-based telescopes at short wavelengths (up to the near-infrared) is not high enough (Strehl ratio of the order of 50\%). Second, the coronagraphic devices offered until a few years ago date back to the original century-old invention by French astronomer Bernard Lyot \citep{Lyot1939}. Paradoxically, the diversity of new coronagraphs developed in the lab during past 15 years is far greater than the number of real science discoveries that they have enabled so far \citep{Guyon2006}. It is also fair to say that, apart from very few exceptions \citep{Riaud2006, Mawet2009, Mawet2011, Serabyn2009, Serabyn2010, Lagrange2009b, Quanz2010, Boccaletti2012}, investments have mainly been made to develop the technologies in the lab rather than actually bring them to the telescope for actual on-sky tests and operations \citep{Mawet2012}, from which so much can be learned while enabling limited but actual science despite the average image quality in the near-infrared.

As a few dedicated experiments have already suggested \citep{Mawet2010,Serabyn2010}, this situation is about to change. Indeed, second-generation instruments start to roll in at major observatories, e.g., the Gemini Planet Imager \citep[GPI,][]{Macintosh2008}, VLT/SPHERE \citep{Beuzit2008,Kasper2012}, and a few others. They promise far greater image quality and stability, enabling new-generation coronagraphs to deliver their best performances. Not to be forgotten are first generation instruments, which still possess untapped potential, that only 10 years or so of operation and understanding allow us to fully unleash \citep{Girard2012,Mawet2012}.

Here we present the successful outcome of 8 years of technological developments to make the so-called annular groove phase mask coronagraph \citep[AGPM,][]{Mawet2005a} a reality ready to be installed at the telescope. The AGPM is an optical vortex made out of a subwavelength grating. The vortex coronagraph is one of the most advanced new-generation coronagraphs and its interest lies in its ability to reach high contrast at very small inner working angles (IWA), while maintaining high throughput over a full $360^\circ$ field of view. This has already been proved in the literature with several successful demonstrations and results, both in the lab and on sky \citep{Mawet2010, Serabyn2010} but at shorter wavelengths (visible and near-infrared). This paper tackles the difficult problem of designing and manufacturing vector vortex coronagraphs for the mid-infrared, where the demand for efficient coronagraphs is increasing following the recent success of high contrast imaging of exoplanets and circumstellar disks in the L band around 4 $\mu$m \citep{Moerchen2007, Lagrange2010, Quanz2010, Quanz2011, Kenworthy2013}. L band is indeed an ideal filter for exoplanet searches on ground-based telescopes:
\begin{itemize}
\item The L-band contrast of planetary-mass companions with respect to their host stars is predicted to be more favorable than at shorter wavelengths \citep{Baraffe2003,Fortney2008, Spiegel2012} so that lower-mass, older objects can be addressed.
\item An additional advantage of L band over shorter wavelengths is the better adaptive optics image quality, with Strehl ratios typically between 70\% and 90\%.
\end{itemize}
These advantages by far compensate the increased sky background in the thermal infrared. In this context, the AGPM will be competitive, for it is designed to reduce the stellar contribution by $\Delta L' \ge$ 7.5 magnitudes at very small IWA (down to $0.9\lambda/D$). An L-band AGPM on an AO-assisted 10-m class telescope should be capable of imaging giant planets at a projected separation of only about 1\,AU from stars located at 10\,pc.

After reviewing the principle of vortex coronagraphs and of the AGPM in \textsection{\ref{sec:vortex}}, we describe in \textsection{\ref{sec:manu}} the design, manufacturing and expected performance of L-band AGPM coronagraphs. In \textsection{\ref{sec:labo}}, we detail the optical setup of the coronagraphic testbench and present the measured performance of our best L-band AGPM. Finally, we conclude by discussing the perspectives of on-sky applications and of future improvements.


\section{The AGPM: a vector vortex coronagraph} \label{sec:vortex}

Optical vortices occur when the phase structure of light is affected by a helical ramp around the optical axis, $e^{\imath l \theta}$, where $\theta$ is the focal plane azimuthal coordinate and $l$ is the vortex topological charge, i.e., the number of times the geometric phase $\phi_{p}$ \citep[][see also below]{Pancharatnam1956} accumulates 2$\pi$ along a closed path $s$ surrounding the phase singularity:
\begin{equation}
l=\frac{1}{2\pi} \oint \! \nabla \phi_{p} \, ds \; .
\label{eq:topo}
\end{equation}
The phase dislocation forces the amplitude to zero at its centre, which is a singularity. Nature indeed prevents the phase from having an infinite number of values at a single point in space, which is non-physical, by simply nulling the light locally. When centred on the diffraction pattern of a star seen through a telescope, optical vortices affect the subsequent propagation by redirecting the on-axis starlight outside the geometric image of the pupil. This diffracted starlight is then blocked when passing through a diaphragm in the so-called Lyot stop (LS) plane, slightly undersized compared to the input pupil. 

The perfect starlight attenuation of an optical vortex coronagraph was proven analytically \citep{Mawet2005a, Foo2005, Jenkins2008} for any non-zero even values of $l$ by applying the result of the Weber-Schafheitlin integral\footnote{This integral reduces to Sonine's for the $l=2$ case.} \citep{Abramowitz1972,Sneddon1951} to the Fourier transform of the product of a vortex ramp phase ($e^{\imath l \theta}$) with an ideal Airy pattern ($\frac{2J_1(k\rho R)}{k\rho R}$), where $R$ is the circular input pupil radius, $k$ the wave number, and $\rho$ the radial coordinate in the focal plane. For instance, with a topological charge $l=2$, the amplitude of the electric field in the Lyot stop plane, with $(r,\gamma)$ coordinates, becomes
\begin{equation}
E_{\rm LS}(r,\gamma) = 
\left\{ \begin{array}{lll}
0 & , \; r < R \\
e^{\imath 2 \gamma} \left(\frac{R}{r}\right)^2 & , \; r \ge R 
\end{array} \right.
\label{eq:ELS}
\end{equation}
The signal is perfectly nulled inside the pupil area, while the redirected light outside the pupil area is blocked by the Lyot stop. Naturally, this occurs only for the on-axis starlight whereas light from off-axis objects misses the centre of the vortex and propagates normally. The vortex coronagraph is known for its near-perfect coronagraphic properties: small IWA (down to $0.9\lambda/D$), high throughput, clear off-axis $360\degr$ discovery space, and simplicity.

An optical vortex can be either scalar or vectorial. Scalar optical vortices are based on a structural phase ramp such as a piece of glass shaped as a helix. The accuracy to be achieved in the glass thickness is a real technological challenge which still remains poorly mastered, even using the most recent micro-fabrication techniques. More importantly, the induced phase profile of scalar vortices is highly dependent on the wavelength of the incoming light. Therefore, these are not suitable for significantly wide spectral bands such as those commonly used as astronomical filters. A vectorial (or vector) vortex coronagraph (VVC), on the other hand, can overcome these limitations. Instead of a structural phase ramp, the VVC is based on a circularly symmetric halfwave plate, i.e., a mask that affects the transverse polarisation state cyclicly around the centre. This creates a geometric phase $\phi_{p}$ known as the Pancharatnam-Berry phase, which is half of the solid angle subtended by the polarisation cycle on the Poincar\'e sphere \citep{Pancharatnam1956, Berry1987}. It was rigorously shown \citep{Mawet2005a,Niv2006} that vectorial vortices present the same phase ramp as scalar vortices.

\begin{figure}[!t]
\begin{center}
\includegraphics[width=8cm]{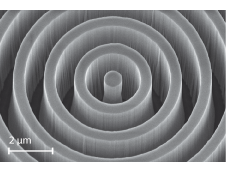}
\end{center}
\caption{Scanning electron microscope (SEM) picture of the centre of an Annular Groove Phase Mask (AGPM) made out of diamond and dedicated to coronagraphic applications in the L band.}
\label{fig:semcenter}
\end{figure}

Different types of VVCs exist. A technology using liquid crystal polymers (LCP-VVC) recently showed excellent results \citep{Serabyn2010, Mawet2010}. Yet, this technology is limited so far to the visible and near-infrared wavelength regimes (up to 2.4 $\mu$m), and is not suitable for the highly coveted mid-infrared region. Another technological route is the Annular Groove Phase Mask (AGPM, see Fig.~\ref{fig:semcenter}) using subwavelength gratings (SG-VVC), which are particularly adapted to longer wavelengths \citep{Mawet2005b, Delacroix2010}. Subwavelength gratings (SGs) are gratings with a period $\Lambda$ smaller than the illuminating wavelength $\lambda$ divided by the refractive index $n$ of the substrate (assuming that the incident light is perpendicular to the grating surface and that the surrounding medium is air). Thanks to their property of form birefringence, which gives two different refractive indices $n_{\rm TE}$ (transverse electric) and $n_{\rm TM}$ (transverse magnetic), such SGs can produce a phase shift between the two polarisation components:
\begin{equation}
\Delta \Phi_{{\rm TE}-{\rm TM}}(\lambda)=\frac{2\pi}{\lambda} h \, \Delta n_{\rm form}(\lambda) \; .
\label{eq:dphi}
\end{equation}
By carefully selecting the grating parameters (substrate and geometry), this phase shift can be made quasi-independent of the wavelength over a wide spectral bandwidth, thereby synthesizing an artificial birefringent achromatic wave plate \citep{Kikuta1997}. We have recently shown \citep{Delacroix2012b} that diamond is a good material for manufacturing achromatic half-wave plates for mid-infrared wavelengths (e.g., L and N bands centred respectively around 3.8 $\mu$m and 10 $\mu$m). Here, we focus on applications in the L band, ranging from 3.5 to 4.1 $\mu$m.


\section{AGPM-L design and manufacturing} \label{sec:manu}

	\subsection{Design} \label{sec:design}

The design of the grating was conducted in complete synergy with the manufacturing process, described in Sect.~\ref{sec:metro}. We performed realistic numerical simulations using the rigorous coupled wave analysis (RCWA), which resolves the Maxwell equations in the frequency domain and gives the entire diffractive characteristics of the studied structure \citep{Moharam1981}. Although the theory states that a perfectly achromatic coronagraph would provide a perfect nulling of the on-axis starlight, in practice, imperfections remain. In this paper, we use the \textit{null depth} metrics instead of the peak-to-peak attenuation. The null depth quantifies the contrast provided by the coronagraph, integrated over the whole point spread function (PSF). As explained further in Sect.~\ref{sec:datproc}, this metrics is well suited to coronagraphy since it takes into account the changes induced by the coronagraph itself to the PSF profile. The theoretical null depth $N_{\rm theo}$ is given by the following wavelength-dependent expression \citep{Mawet2005a}:
\begin{equation}
N_{\rm theo}(\lambda)=\frac{I_{\rm coro}(\lambda)}{I_{\rm off}(\lambda)}=\frac{[1-\sqrt{q(\lambda)}]^2+\epsilon(\lambda)^2\sqrt{q(\lambda)}}{[1+	\sqrt{q(\lambda)}]^2} \; ,
\label{eq:null}
\end{equation}
where $I_{\rm coro}$ (resp.\ $I_{\rm off}$) is the signal intensity when the input beam is centred on (resp.\ off) the coronagraph, $\epsilon(\lambda)$ is the phase error with respect to $\pi$, and $q(\lambda)$ is the flux ratio between the two polarisation components, TE and TM. Note that in the theoretical case where the coronagraphic profile remains identical to the original PSF profile, this formula of the null depth is also valid for the peak-to-peak attenuation, which was demonstrated in an ideal case by \citet{Mawet2005a}.

\begin{figure}[!t]
\begin{center}
\includegraphics[width=8cm]{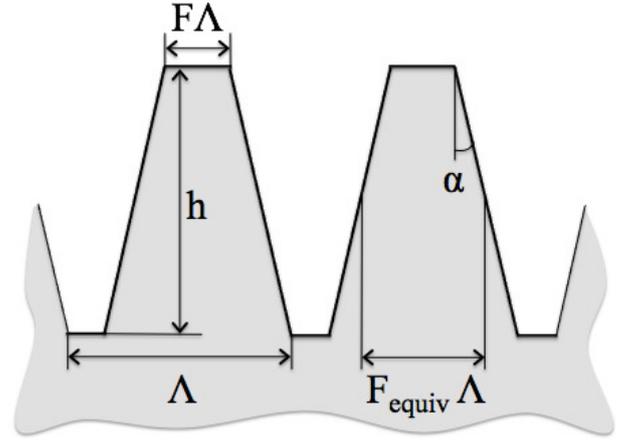}
\end{center}
\caption{Schematic diagram of a trapezoidal grating. The grating sidewalls have an angle $\alpha$ and an average width $F_{equiv}\Lambda$.}
\label{fig:grat}
\end{figure}

\begin{figure}[!t]
\begin{center}
\includegraphics[width=8cm]{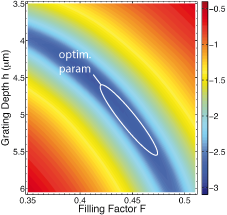}
\end{center}
\caption{RCWA multi-parametric simulation: mean null depth (logarithmic scale) over the whole L band (3.5-4.1\,$\mu$m) with $\alpha$ ranging from $2.7\degr$ to $3.2\degr$. The period is set to $\Lambda=1.42 \, \mu$m (SG limit).}
\label{fig:multi}
\end{figure}

Eq.~\ref{eq:null} involves all the geometrical parameters of the grating (period $\Lambda$, filling factor $F$, depth $h$, sidewall angle $\alpha$), illustrated in Fig.~\ref{fig:grat}. The period $\Lambda$ is the only fixed parameter in our design. It is determined by the subwavelength limit ($\lambda/n$). Considering a little margin for the spectral band lower bound ($\simeq 3.4\, \mu$m), and considering the refractive index of diamond ($\simeq 2.38$), the period should be $\leqslant 1.42\, \mu$m. The other parameters of the manufactured grating are free parameters whose values are provided by numerical optimisation. Some of the parameters, like the depth $h$, the filling factor $F$ and the sidewall angle $\alpha$, may vary slightly during the etch process. The only way to measure them precisely is by cracking the sample along a diameter. By doing so, we measured that the angle $\alpha$ is usually $2.95\degr \pm 0.25\degr$. Due to the uncertainty on the slope, we sought a robust design which performs well even slightly outside of the optimal parameter set. In order to find the target design parameters, we computed two-dimensional maps of the theoretical null depth as a function of the filling factor $F$ and of the depth $h$, for several values of the angle $\alpha$ ranging from $2.7\degr$ to $3.2\degr$. The mean of all these maps (see Fig.~\ref{fig:multi}) gives the targeted specifications, i.e.,~the optimal parameters set region: period $\simeq 1.42\, \mu$m, filling factor $\simeq 0.45$, depth $\simeq 5.2\, \mu$m, and sidewall angle $\simeq 2.95 \degr$. For these optimal parameters, the mean null depth over the whole L band (3.5-4.1\,$\mu$m) equals $5\times10^{-4}$.

	\subsection{Manufacturing and metrology} \label{sec:metro}

The AGPM pattern is etched onto a commercial optical-grade polycrystalline diamond window of 10\,mm diameter and 300\,$\mu$m thick, grown by chemical vapour deposition (Element Six Ltd.). The etching of the grooves involves techniques inherited from the micro-electronics industry. Electron-beam lithography was used to produce the pattern on a silicon wafer. Metal masking layers were deposited on the diamond by sputtering, and nano-imprint lithography was used to transfer the pattern to the top of the mask. Finally, reactive ion etching was used to etch the pattern first into the metal mask and then into the diamond. A similar manufacturing process has been described before \citep{Karlsson2010,Delacroix2010}. This process was painstakingly optimised to achieve the high pattern homogeneity, precision and aspect ratio necessary for half-wave plates and AGPMs in the N band \citep{Delacroix2012a}. This included introducing a soft silicone stamp in the nano-imprint process and adding masking layers to better control the mask thickness and profile. 

\begin{figure}[!t]
\begin{center}
\includegraphics[width=8cm]{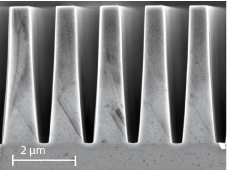}
\end{center}
\caption{SEM picture of the cleaved subwavelength grating, from which are deduced the geometric parameters of the AGPM-L4 profile: line width $\simeq$0.58 $\mu$m (i.e., filling factor $\simeq$0.41) and depth $\simeq$4.7 $\mu$m.}
\label{fig:semSG}
\end{figure}

For the L-band AGPM only minor adjustments had to be made to take into account the difficulties of etching a finer grating. In particular, the etch rate and sidewall angle were both more sensitive to variations in groove width and the depth of the grooves was even more difficult to measure with sufficient precision. In order to make precise measurements of the profile and depth, the sample had to be cracked and the cross-section imaged by scanning electron microscopy (SEM), as seen in Fig.~\ref{fig:semSG}. In the N-band case, the depth could be estimated with reasonable precision without cracking the sample or removing the mask, so etching could be carried out in steps with depth measurements in between until the correct depth was reached. For the L-band, this was not possible and a slightly different approach had to be taken. After etching the metal mask, the sample was imaged by SEM in order to measure the line width. This width did not change noticeably during diamond etching, so the top width ($F\Lambda$~in Fig. \ref{fig:grat}) of the final structure was known beforehand. Test samples with identical masks as the AGPM sample were then etched and cracked in order to determine the etch rate and sidewall angle. Two samples etched to different depths close to the final depth were required, as the etch rate slows when the grooves get deeper. The optimal depth was then recalculated by RCWA with the measured top width and sidewall angle fixed. Finally, the AGPM was etched for an appropriate length of time to achieve the desired depth. A detailed account of the masking and etching process used both for the N band and the L band will appear elsewhere (Forsberg et al., 2013, submitted to Diamond and Related Materials). The grating depth of the finished components was confirmed by careful parallax measurements between SEM pictures at five different tilt angles.

\begin{table}[!t]
\caption{Measured profile parameters and simulated raw null depth for the four manufactured L-band AGPMs, and comparison with the optimal design.}
\label{tab:param}
\centering
\begin{tabular}{cccccc}
\hline \hline
name & period $\Lambda$ & fill $F$ & depth $h$ & angle $\alpha$ & $N_{\rm theo}$ 
\\ \hline 
AGPM-L1 & 1.42 $\mu$m & 0.35 & 4.2 $\mu$m & 2.40$\degr$ & 0.0267 \\ 
AGPM-L2 & 1.42 $\mu$m & 0.36 & 3.6 $\mu$m & 2.65$\degr$ & 0.0116 \\ 
AGPM-L3 & 1.42 $\mu$m & 0.44 & 5.8 $\mu$m & 3.25$\degr$ & 0.0048 \\ 
AGPM-L4 & 1.42 $\mu$m & 0.41 & 4.7 $\mu$m & 3.10$\degr$ & 0.0010 \\ 
\hline
optimum &  1.42 $\mu$m & 0.45 & 5.2 $\mu$m & 2.95$\degr$ & 0.0005 \\ 
\hline
\end{tabular}
\end{table}

Four finished components have been made to date. The first two, AGPM-L1 and AGPM-L2, were initial tests. These were etched with the exact same methods as N-band components. Further, the e-beam-written master used in the patterning of these two had too narrow lines. AGPM-L3 was an experiment with a different etch recipe giving a grating profile of trapezoidal walls with a triangular tops, designed to reduce the problem of ghost reflections (see Sect. \ref{sec:eperf}). AGPM-L4 finally, was produced using the method described above to achieve a close to optimal depth for its actual fill factor and sidewall angle. That we were able to produce nearly optimal grating parameters with so few trials directly reflects the maturity of the fabrication process. The grating parameters of all four components and their simulated raw null depth can be found in Table~\ref{tab:param} (with AGPM-L3 profile approximated as trapezoidal). In the rest of this paper, only the best component, AGPM-L4, will be discussed.


\subsection{Antireflective grating} \label{sec:arg}

\begin{figure}[!t]
\begin{center}
\includegraphics[width=8cm]{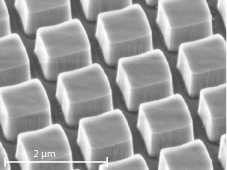}
\end{center}
\caption{SEM picture of the antireflective grating (ARG) etched on the backside of the AGPM.}
\label{fig:semAR}
\end{figure}

\begin{figure}[!t]
\begin{center}
\includegraphics[width=8cm]{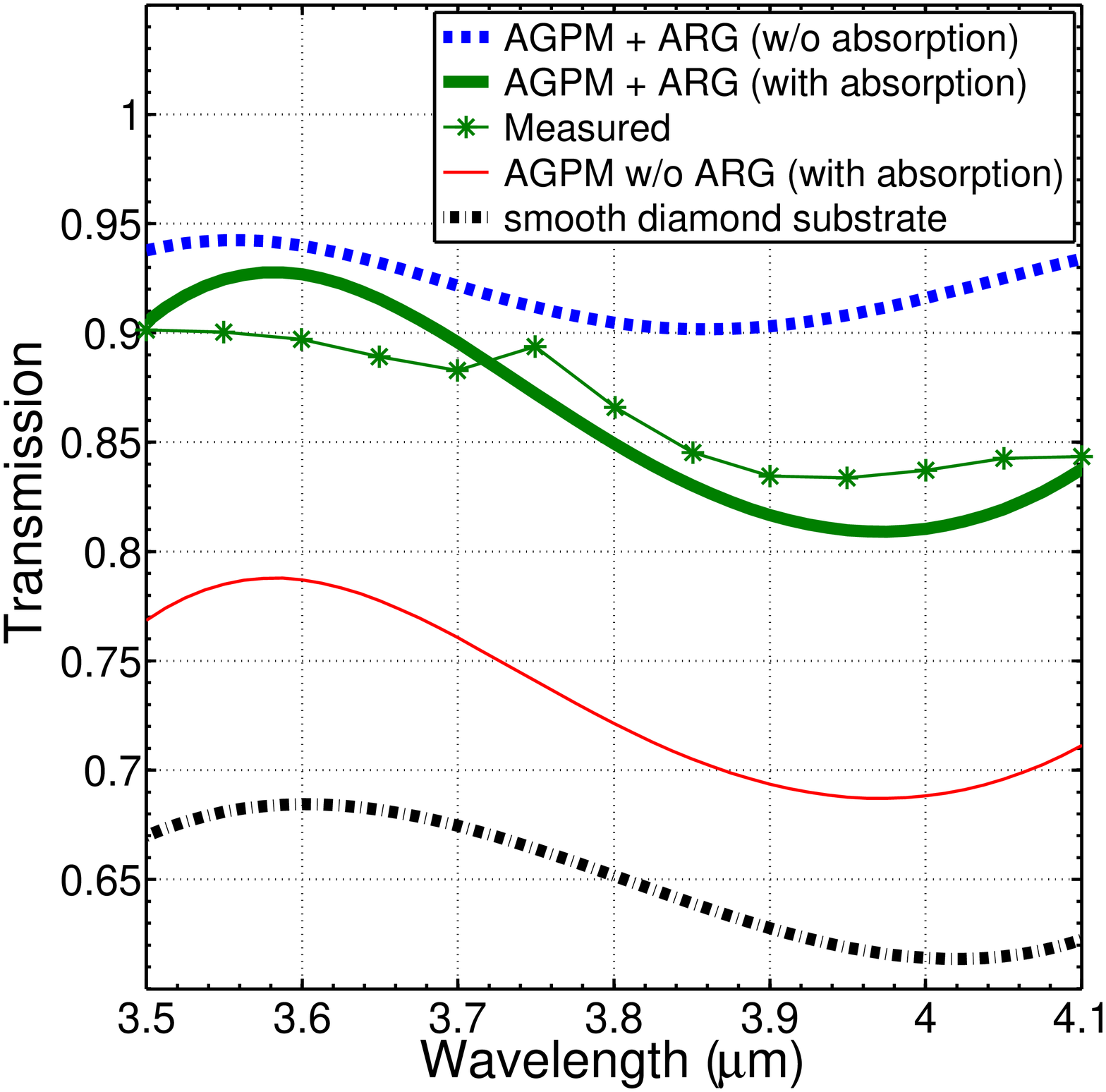}
\end{center}
\caption{L-band transmission measurements of AGPM-L4 with an antireflective grating (ARG) etched on its backside. The figure also shows the calculated transmission curves, with/without the absorption, with/without the ARG, and for a smooth diamond substrate.}
\label{fig:trans}
\end{figure}

In order to avoid incoherent reflections on the internal sides of the AGPM, which would not be cancelled by the optical vortex, an anti-reflective grating (ARG) was etched on the backside of the component, using a very similar diamond etching technique \citep{Karlsson2003} with a binary square shaped structure as can be seen in Fig.~\ref{fig:semAR}. The backside reflection, which is high for diamond in the L band ($\sim 17\%$), is significantly reduced ($\sim 1.9\%$) thanks to the use of the ARG. The total transmission of the AGPM was measured on a Perkin-Elmer 983 spectrophotometer and compared to the theoretical transmission calculated with RCWA (Fig.~\ref{fig:trans}). The absorption, which is significant in the L band (especially around 4 $\mu$m), was taken into account in the calculation. The result is very satisfying, with an average total transmission $\sim 87\%$.

\subsection{Expected performance} \label{sec:eperf}

\begin{figure}[!t]
\begin{center}
\includegraphics[width=8cm]{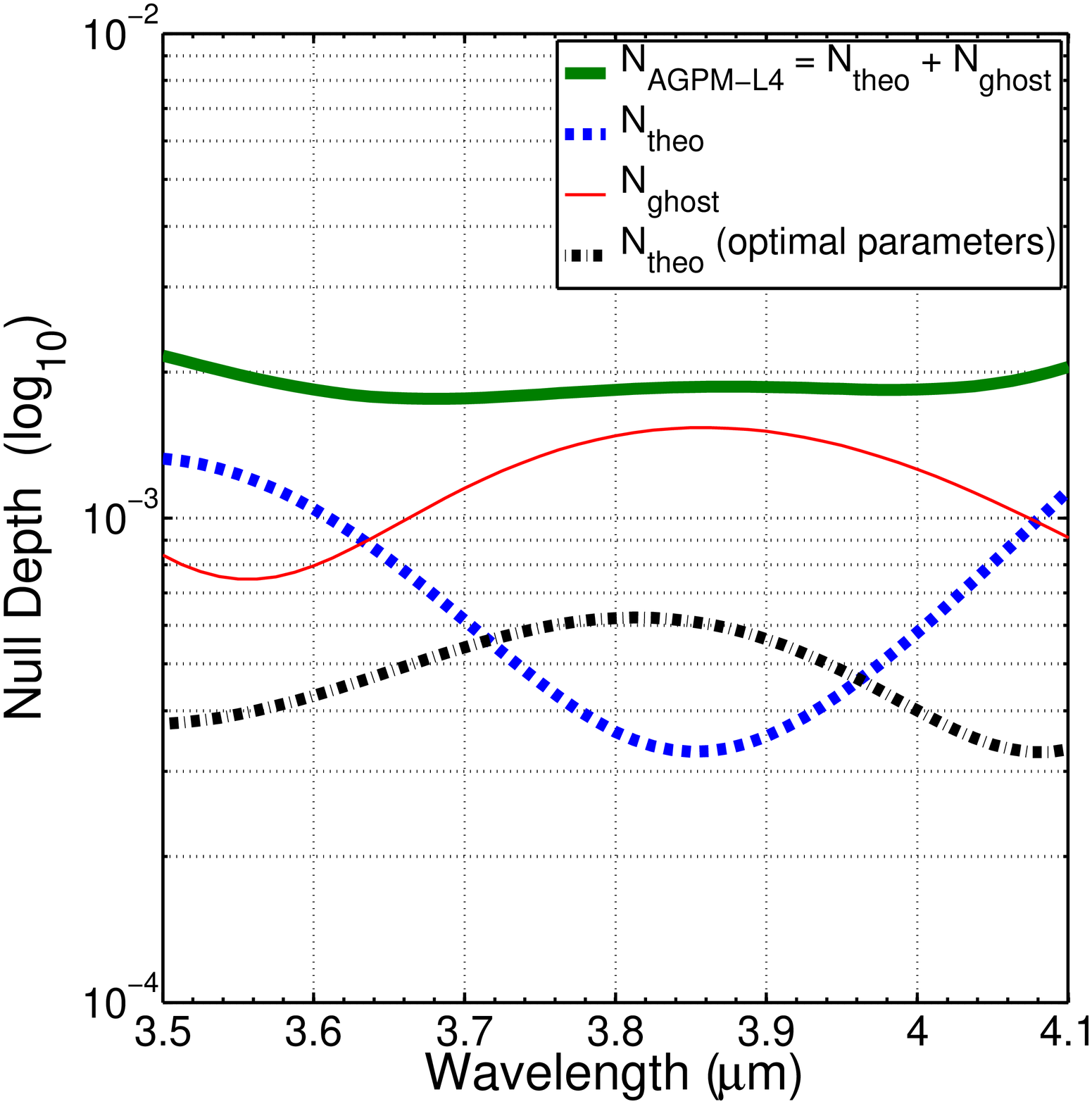}
\end{center}
\caption{L-band expected performance of AGPM-L4, calculated as the sum of the theoretical raw null depth $N_{\rm theo}$ and the ghost contribution $N_{\rm ghost}$. The figure also shows the theoretical raw null depth obtained with the optimal design presented in Table~\ref{tab:param}.}
\label{fig:null}
\end{figure}

The performance of the coronagraph, i.e., the measured raw null depth $N_{\rm AGPM}$, can be described as the sum of two terms: 
\begin{equation}
N_{\rm AGPM}=N_{\rm theo} + N_{\rm ghost} \; .
\label{eq:Nagpm}
\end{equation}
The first term is the theoretical null depth (Eq.~\ref{eq:null}), which is limited by the manufacturing accuracy. The second term is induced by the ghost, i.e., the unwanted internal double reflection described in Sect.~\ref{sec:arg}. The ghost contribution $N_{\rm ghost}$ is the ratio between the ghost intensity and the signal intensity when the input beam is centred off the coronagraph:
\begin{equation}
N_{\rm ghost}=\frac {I_{\rm ghost}} {I_{\rm off}}=\frac{T_{\rm SG}  T_{\rm ARG} R_{\rm ARG}  R_{\rm SG} }{T_{\rm SG} T_{\rm ARG}  } = R_{\rm ARG}  R_{\rm SG} \; ,
\label{eq:Nghost}
\end{equation}
where $T_{\rm SG} $, $T_{\rm ARG}$, $R_{\rm SG}$ and $R_{\rm ARG}$ are the transmit-\linebreak[4]tances and reflectances for both the subwavelength grating (SG) and anti-reflective grating (ARG) interfaces. In our case $R_{\rm ARG} R_{\rm SG} \simeq 10^{-3}$, which is quite small but not negligible in the case of the best component (AGPM-L4), whose theoretical null depth is also $10^{-3}$ (see Table~\ref{tab:param}). The expected performance of AGPM-L4 was calculated and is presented in Fig.~\ref{fig:null}. The mean null depth over the whole L band (3.5-4.1 $\mu$m) is $2\times10^{-3}$.

\section{Laboratory results} \label{sec:labo}

After the manufacturing of different AGPMs with slightly varying grating heights and filling factors, coronagraphic tests were carried out in the L band at the Observatoire de Meudon, on the same optical bench (YACADIRE) as the one used for characterising the VLT-SPHERE coronagraphs \citep{Boccaletti2008}.


\subsection{Description of the coronagraphic test bench} \label{sec:bench}

\begin{figure}[!t]
\begin{center}
\includegraphics[width=8cm]{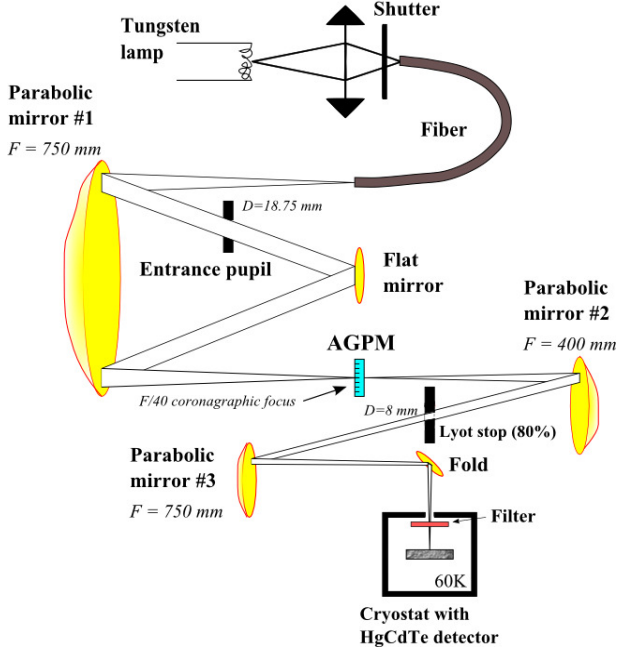}
\end{center}
\caption{Optical layout of the YACADIRE coronagraphic test bench.}
\label{fig:bench}
\end{figure}

\begin{figure}[!t]
\begin{center}
\includegraphics[width=8cm]{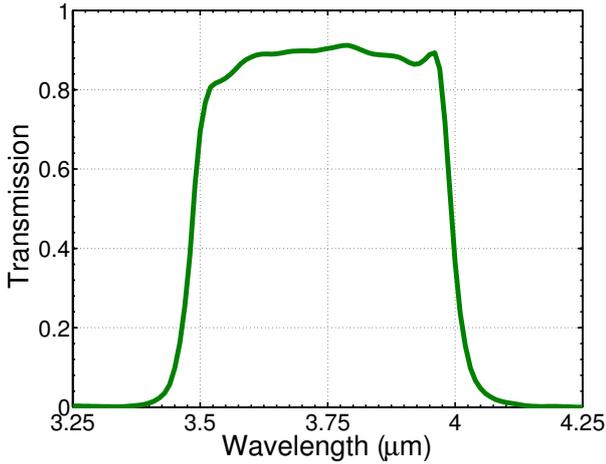}
\end{center}
\caption{Transmission curve measured at cold for the wide L-band filter (Spectrogon BBP-3500-4000nm).}
\label{fig:filter}
\end{figure}

The setup of the coronagraphic test bench, taking into account some minor modifications, is represented in Fig.~\ref{fig:bench}. The bench is approximately 1.5\,m long and located in a clean room. A tungsten lamp is used to feed a single-mode fibre (our star simulator) transmitting wavelengths up to 5\,$\mu$m. Gold coated parabolic and flat mirrors are placed successively in the light path between the fibre and the detector, so that the beam alternates from pupil to focal planes. This choice of a complete set of reflective optical components avoids inherent chromatic aberrations. The AGPM to be tested is placed in an intermediate focal plane between two parabolic mirrors. With a focal distance of $F=750$\,mm for the first parabola and a beam diameter of 18.75\,mm, the f-number of the beam impinging on the AGPM is $F/40$. The core of the Airy spot created on the AGPM is $F\lambda/D\simeq 150 \, \mu$m wide. A Lyot stop blocking the diffracted light is then located after the second parabolic mirror, reducing the pupil to 80\% of its diameter. The beam is subsequently focused at $F/94$ with the third parabolic mirror on a HgCdTe infrared detector placed in a cryostat cooled down to 60\,K. Inside the cryostat, the beam first passes through a wide L-band filter (3.5-4.0\,$\mu$m), whose transmission characteristics are given in Fig.~\ref{fig:filter}.

	 \subsection{Working and measurement principles}\label{sec:workprinc}

We started our coronagraphic tests by placing the AGPM precisely at the focus of the beam. A coarse positioning was achieved using a visible laser, by looking at the diffraction rings created upon propagation through the AGPM grating. The fine positioning was then performed in the L band, by minimising the flux reaching the camera. A second figure of merit in this optimisation was the shape of the coronagraphic PSF, which is expected to be symmetric when the AGPM is properly centred in $x$ and $y$ (the horizontal and vertical axes). The adjustment of the focus ($z$~axis) was then based solely on the quality of the on-axis light extinction.

Series of 100 frames of 40\,ms were recorded with the AGPM placed at the optimal position (referred hereafter as the \emph{coronagraphic} frames) and subsequently with the AGPM far from this position (typically at 1 mm) to measure a reference PSF without coronagraphic effect, but with the beam still propagating through the diamond substrate (referred hereafter as the \emph{off-axis} frames). These two sorts of frames (coronagraphic and off-axis) were used to evaluate the null depth as described hereafter (Sect. \ref{sec:datproc}). Series of 100 frames were also recorded for various offsets of the AGPM from the optimal position, both in $x$, $y$ and $z$. In each case, 10 to 20 background frames were recorded to allow a basic treatment of our images. All exposure times were limited to 40\,ms due to the saturation of the thermal background.

	\subsection{Data processing}\label{sec:datproc}

The first step consisted in computing the median of the background images, before subtracting it from either the coronagraphic or the off-axis frames. As we noticed that the background was sometimes slightly over-subtracted (resulting in negative values in the reduced images), a scaling of the background was implemented by using the ratio of the mean intensity in identical rings in both the coronagraphic (or off-axis) and the background frames. A flat field was measured and subsequently divided from the background-subtracted images. Noting $f_{\rm scal}$ the aforementioned scaling factor, the reduced images $I_{{\rm red},j}$ are given by:
\begin{equation}
\label{eq:BckgScaling}
I_{{\rm red},j} = \frac{I_{j} - f_{{\rm scal}}I_{B,{\rm med}}}{i_F} ~~~~\mathrm{with}~~~f_{\rm scal} = \frac{\left\langle {\rm ring}(I_{\rm coro})\right\rangle}{\left\langle{\rm ring}(I_{B,{\rm med}})\right\rangle} \; ,
\end{equation}
where $I_j$ is the $j^{\rm th}$ image within the coronagraphic or off-axis frames, $I_{B,{\rm med}}$ the median of the background images, $i_F$ the normalised flat field, and $\langle\rangle$ represents the average over a number of pixels. Next, bad pixels and cosmic rays were adequately eliminated. Finally, all the reduced images were stacked to increase the signal to noise ratio of both the off-axis and coronagraphic images. Aperture photometry was then carried out on these final images in order to measure the flux in the coronagraphic and off-axis PSF, and thereby estimate the performance of our AGPM across the full L band. 

Although the peak-to-peak attenuation is a widely used metrics to quantify the coronagraphic performance, the presence of the AGPM slightly changes the PSF profile near the axis. This phenomenon is characteristic of all coronagraphs. Therefore, a more robust metric needs to be used, taking into account both the coronagraphic and the off-axis frames mentioned in Sect.~\ref{sec:workprinc}. In our case, the figure of merit is the raw null depth, defined as the ratio between the integrated flux over a certain area around the center of the final coronagraphic image and the integrated flux over the same area in the off-axis image. The question of the relevant size of this area is interesting. Using the full images would lead to a pessimistic result as it would integrate a lot of background and high frequency artifacts. Using the sole central peak seems to be more appropriate. As we wish to compare our results with the theoretical predictions of \citet{Riaud2010}, we decided to use an area of radius equal to the FWHM of the PSF ($\sim \lambda/D$). This generally results in a good agreement with the theoretical formula (Eq.~\ref{eq:Nagpm}), because most of the intensity in the image is contained within the FWHM. Our raw null depth is thus defined as:
\begin{equation}
N_{\rm AGPM} = \frac{\int_0^{\rm FWHM} \int_0^{2\pi} \tilde{I}_{\rm coro}(r,\theta) \: r \: \ud r \: \ud\theta}{\int_0^{\rm FWHM} \int_0^{2\pi} \tilde{I}_{\rm off}(r,\theta)  \: r \: \ud r \: \ud\theta} 
\end{equation}
where $\tilde{I}_{\rm off}$ and $\tilde{I}_{\rm coro}$ are the medians of the reduced off-axis and coronagraphic images, respectively. This definition would be perfectly equivalent to Eq.~\ref{eq:Nagpm} if the coronagraphic profile was identical to the original PSF profile.

	\subsection{Measured performance and comparison with theory} \label{sub:Perfos}

\begin{figure}[!t]
\begin{center}
\includegraphics[width=8cm]{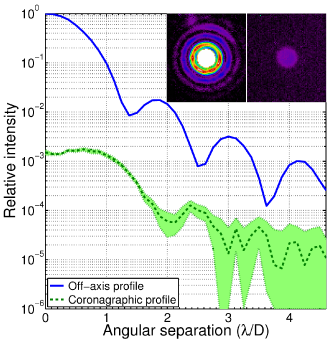}
\end{center}
\caption{Azimuthally averaged PSF (plain blue curve) and coronagraphic profiles (dash green curve) with confidence interval (green shaded region). The top-right inset shows the off-axis (left) and coronagraphic (right) images on a common linear scale. The off-axis image is thresholded and the coronagraphic image underexposed, conveying the true raw dynamic range achieved by the coronagraph, with a measured raw null depth of $2\times10^{-3}$ and a raw contrast of $6 \times 10^{-5}$ measured at $2\lambda/D$.}
\label{fig:prof}
\end{figure}

The raw null depth measured by the described procedure was $2\times10^{-3}$ over the L~band (or alternatively, a rejection ratio $R=1/N\simeq 500$), which is perfectly in line with the expected performance based on our RCWA modeling (see Sect.~\ref{sec:eperf}). In Fig.~\ref{fig:prof}, the measured radial profile for the coronagraphic PSF (dotted line) is compared to the off-axis PSF profile (solid line). These profiles were obtained by computing the azimuthal medians in concentric rings of various sizes in our images, with a subsequent normalisation to the peak value of the off-axis PSF. We computed a confidence interval on our median coronagraphic profile by measuring a series of 45 coronagraphic profiles on an integration time of about 40\,sec each (i.e., 1000 frames), and by computing the standard deviation of the profiles in each individual ring. The main source of fluctuations in the coronagraphic profile level is associated to background instabilities. The green shaded region in Fig.~\ref{fig:prof} shows that the coronagraphic profile is measured with a sufficient signal-to-noise ratio up to about $3\lambda/D$. Beyond this point, the intensity in the median coronagraphic profile is dominated by background subtraction residuals. The raw contrast delivered by the AGPM at $2\lambda/D$ equals $6 \times 10^{-5}$, which corresponds to 10.5\,mag. Beyond $3\lambda/D$, the contrast quickly falls below $10^{-5}$ (i.e., 12.5\,mag) and cannot be measured any more.

\begin{figure}[!t]
\begin{center}
\includegraphics[width=8cm]{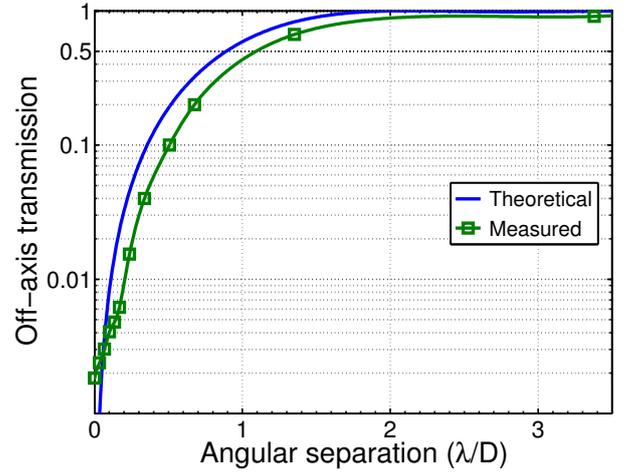}
\end{center}
\caption{Theoretical and measured off-axis transmission as a function of the angular separation between the beam and AGPM centre, expressed in resolution elements ($\lambda/D$).}
\label{fig:off}
\end{figure}

In a second phase, we explored the dependency of the coronagraphic attenuation as a function of the offset of the AGPM with respect to the optimal position. The results of these measurements are shown in Fig.~\ref{fig:off}. The agreement with the theoretical off-axis transmission profile of \citet{Riaud2010} is reasonable, which further suggests that our AGPMs produce the textbook vortex effect. Finally, we measured the sensitivity of the null depth to the focus. The null depth was divided by a factor 2 for a defocus of about $\pm1.5$\,mm at $F/40$, translating to a defocus aberration of $\pm120$\,nm peak-to-valley ($\sim30$\,nm RMS).


\section{Conclusions and perspectives} \label{sec:conclu}

In this paper, we have presented the design, manufacturing and measured performance of the first vortex phase mask coronagraphs based on the use of subwavelength gratings. Our Annular Groove Phase Masks, etched on diamond substrates, produce on-axis light rejection reaching a factor up to 500 across the L band (3.5-4.0\,$\mu$m), which represents the best broadband performance of any mid-infrared phase mask to our knowledge. Assuming a high-quality input wave front, the use of our L-band AGPM on a telescope would result in an achievable raw contrast ranging from about $10^{-3}$ (7.5 mag) at an angular separation of $1\lambda/D$ to about $6 \times 10^{-5}$ (10.5~mag) at $2\lambda/D$, and $10^{-5}$ (12.5~mag) at $3\lambda/D$. This underlines the great potential of AGPM-L for exoplanet detection and characterisation at small angular separation from bright nearby stars. 

In November 2012, AGPM-L3 was installed at the focus of NAOS-CONICA, the infrared adaptive-optics camera of the VLT \citep{Lenzen2003,Rousset2003}. Although the NACO image quality is not perfect in the L band, with a Strehl ratio around $70-85\%$, the very first on-sky tests suggest that the AGPM coronagraphic mode lives up to our expectations. These recent results will be presented in a forthcoming paper \citep{Mawet2013}. The installation of AGPM-L4 on LMIRCam, the L-band camera of the LBTI \citep{Skrutskie2010}, is currently on-going. Thanks to the very high Strehl ratio delivered by the LBT adaptive optics at L band \citep[$> 95\%$,][]{Esposito2011}, we expect that AGPM-L4 will unleash its full potential on this instrument.

There are currently two main limitations to the performance of our L-band AGPMs. The first one is the presence of a ghost, reflected back and forth between the parallel faces of the diamond substrate. To reach starlight extinction better than 1000, a more aggressive antireflective solution will need to be used, based e.g. on an improved subwavelength structure or on a combination of the current design with an appropriate coating. The second limitation resides in the control of the grating parameters during manufacturing, including the slope of the grating walls. RCWA simulations show that, if the grating parameters can be made perfect, a theoretical on-axis attenuation of almost $10^{-4}$ could be reached. It is however expected that for such deep extinction, local phenomena not taken into account by RCWA simulations could become dominant. Such considerations will be fully investigated when designing and manufacturing second-generation AGPMs, for applications on future extremely large telescopes.



\begin{acknowledgements}

The first author is grateful to the financial support of the Belgian Fonds de la Recherche Scientifique (FRIA) and Fonds de solidarit\'e ULg. SH and JS acknowledge support from the Belgian FRS-FNRS FRFC. We also gratefully acknowledge financial support from the Swedish Diamond Center (financed by Uppsala University), and the Communaut\'e fran\c caise de Belgique -- Actions de recherche concert\'ees -- Acad\'emie universitaire Wallonie-Europe. 

\end{acknowledgements}


\bibliographystyle{aa} 
\bibliography{Delacroix13_AGPM} 

\end{document}